\documentclass[12pt]{joptics}
\usepackage{xcolor}

\title{Adding a spin to Kerker's condition: arbitrary scattering direction tuning of nano-antennas with designed excitation}

\author[*]{Lei Wei}
\author{N. Bhattacharya}
\author{H. Paul Urbach}

\affil{Department of Imaging Physics, Delft University of Technology, Lorentzweg 1, 2628CJ,  Delft, The Netherlands}

\affil[*]{Corresponding author: l.wei-11@tudelft.nl}

\dates{Compiled \today}

\begin{document}

\maketitle
\abstract{We describe a method to control the directional scattering of a high index dielectric nanosphere, which utilizes the unique focusing properties of an azimuthally polarized phase vortex and a radially polarized beam to independently excite insde the nanosphere a spinning magnetic dipole and a linearly polarized electric dipole mode normal to the magnetic dipole. We show that by simply adjusting the phase and amplitude of the field on the exit pupil of the optical system, the scattering of the nanosphere can be tuned to any direction within a plane and the method works over a broad wavelength range.}

\section{Introduction}
How to precisely control the scattering properties of subwavelength nanostructures is becoming a more and more important research topic with applications for directional coupling of light into photonic chips\cite{A.V.Zayats_S2013, Xi_JO2014}, displacement metrology\cite{Xi_PRL2016,Neugebauer_NC2016}, directional excitation of quantum emitters\cite{Curto_S2014}, etc.. High index dielectric nanoparticles have not only electric modes but also magnetic modes\cite{Kuznetsov_SR2012}, making them low-loss and more functional alternatives to metallic nanostructures as building blocks for nano-antennas. The interference of the magnetic and electric dipole modes of a dielectric nanoparticle under plane wave illumination can lead to zero backward scattering\cite{FU_NC2013}, similar to what was theoretically predicted by Kerker et al. \cite{Kerker1983}. However, for a given refractive index and radius of the particle, this can only be realized at a single wavelength due to the relation between electric and magnetic field of a plane wave. Independent control of linearly polarized magnetic and electric dipoles can be realized with an excitation configuration composed of orthogonally oriented focused azimuthally and radially polarized beams\cite{Xi_OL2016}. As a result, directional scattering over a broad wavelength range can be achieved. However, in this case two focusing optical systems are needed and because the two dipoles are both linearly polarized, directional scattering is realized only along the direction normal to the plane parallel to the orientation of the electric and magnetic dipoles. Recently, spin-controlled directional coupling of light has drawn lots of attentions and is considered as an application of the spin-orbit interaction of light\cite{KYBliokh2015}. Several experiments have demonstrated the spin controlled directional excitation of surface plasmon and waveguide modes, utilizing the near-field interference of a rotating dipole\cite{A.V.Zayats_S2013, Xi_JO2014}, the transverse spin of the evanescent waves\cite{Connor2014, Petersen2014} or metasurfaces\cite{lins2013}. However, the propagation direction of light controlled by spin is binary corresponding to either left or right circularly polarized light. In the current paper, a method is proposed to overcome this limitation and tune the scattering of a dielectric nanosphere to any direction within a plane. The approach relies on the independent excitation of a spinning magnetic dipole and a linearly polarized electric dipole mode which is oriented normal to the plane of the spinning magnetic dipole. With help of the unique focusing properties of vector vortex beams, we are able to realize such excitation and achieve arbitrary scattering direction tuning in a plane over a relatively broad wavelength range by simply adjusting the amplitude and phase of the beam.\\

\section{Basic physical model}
\begin{figure}[!h]
\centering
{\includegraphics[width=7.8cm]{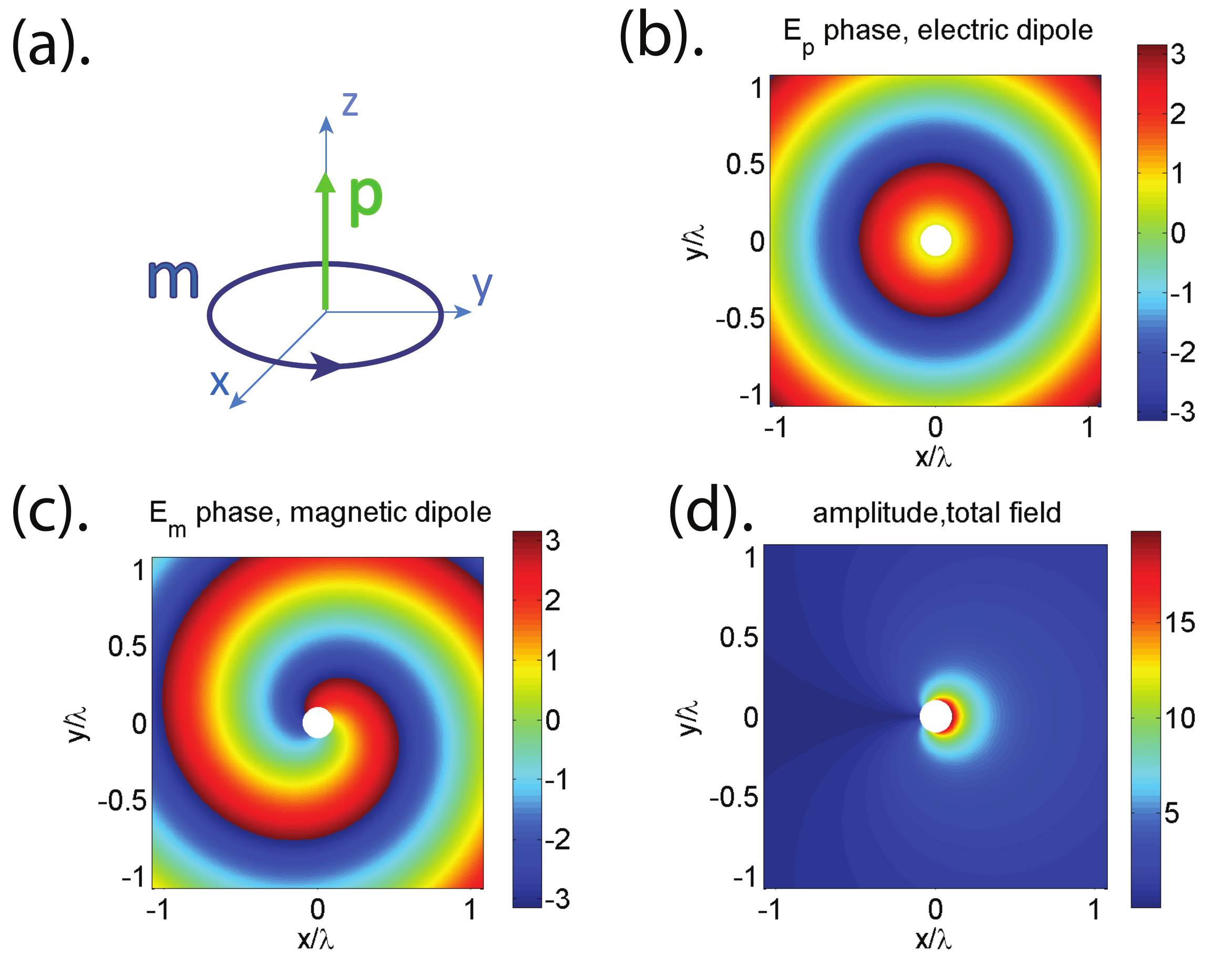}}
\caption{(a).Illustration of a spinning magnetic dipole $\mathbf{m}$ in $xy$ plane and a electric dipole $\mathbf{p}$ along $z$ axis; (b).Radiation electric field in $xy$ plane of the electric dipole is cylindrically symmetric to the $z$ axis; (c).Radiation electric field in $xy$ plane of the spinning magnetic dipole has a spiral phase; (d). Amplitude of the total electric field in $xy$ plane as a result of the two dipole interference.}
\label{fig_spin_dipole}
\end{figure}
Unlike previous approaches based on the interference of linearly polarized magnetic and electric dipoles, we propose a method to actively control the scattering in any direction in the $xy$ plane using the interference of a spinning(circulary polarized) magnetic dipole $\mathbf{m}=m_0(\hat{x}+i\hat{y})$ and an electric dipole $\mathbf{p}=p_0\hat{z}$ linearly polarized in the $z$ direction as illustrated in Fig.\ref{fig_spin_dipole}. It is easy to show\cite{ed_griffith} that the farfield electrical field in the $xy$ plane of the electrical dipole is polarized along the $z$ direction and cylindrically symmetric around the $z$ axis: 
\begin{equation}
\mathbf{E}_p=\hat{z}\frac{p_0k^2}{\epsilon_0}\frac{e^{ikr}}{4\pi r}=\hat{z}E_{zp}\frac{e^{ikr}}{4\pi r},
\label{eq_ed}
\end{equation}
whereas the farfield electric field in the $xy$ plane of the spinning magnetic dipole is also polarized along the $z$ direction and has a spiral phase: 
\begin{equation}
\mathbf{E}_m=\hat{z}(-i)m_0k^2Z_0 e^{i\phi}\frac{e^{ikr}}{4\pi r}=\hat{z}E_{zm}e^{i\phi}\frac{e^{ikr}}{4\pi r},
\label{eq_md}
\end{equation}
where $(r, \phi, \theta)$ is the spherical coordinate, $\epsilon_0$ is the vacuum permittivity, $Z_0=120\pi$ is the impedance of free space and $k=2\pi/\lambda$ is the wavevector in air. If the magnetic and electric dipole are modulated in such a way that $E_{zp}=E_{zm} e^{i\phi_0}$, the total electric field in the $xy$ plane is:
\begin{equation}
E_z=E_{zm} e^{i\phi_0}\frac{e^{ikr}}{4\pi r}+E_{zm}e^{i\phi}\frac{e^{ikr}}{4\pi r}.
\end{equation}
As a result of the spiral phase of $\mathbf{E}_{m}$, the symmetry of the radiation field is broken and there is always a direction $\phi=\phi_0$ where a destructive interference occurs, while in the opposite direction $\phi=\phi_0-\pi$ a constructive interference is formed. By choosing the phase $\phi_0$ of the electric dipole appropriately, the destructive and constructive interference can be tuned to any direction in the $xy$ plane.\\
\section{Proposed realisation}
\begin{figure}[!h]
\centering
{\includegraphics[width=7.8cm]{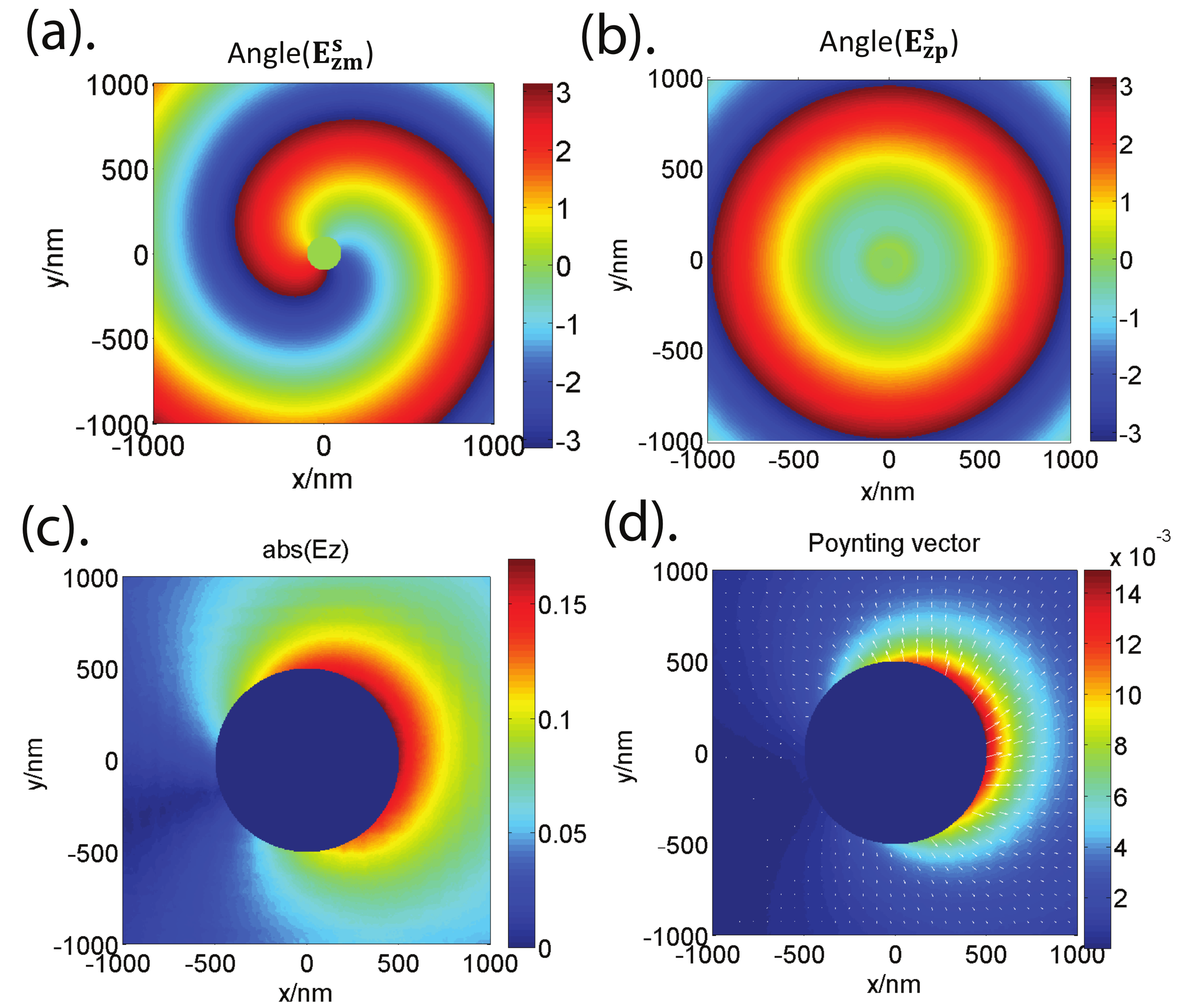}}
\caption{Scattering by a sphere with radius 100 nm and refractive index 5 and for wavelength $\lambda=1028$ nm. (a).Phase of the scattered electric field in the $xy$ plane of the spinning magnetic dipole of the sphere which is excited by the focal field of $\mathbf{E}_h$; (b).Phase of the scattering electric field in the $xy$ plane with the longitudinal electric dipole of the sphere which is excited by the focal field of $\mathbf{E}_e$; (c). Amplitude of the scattered electric field in the $xy$ plane with both dipole excitations using pupil field $\mathbf{E}_h+\mathbf{E}_e$ when $120\pi|b_1\mathbf{H}_f|=|a_1\mathbf{E}_f|$ is fullfilled; (d). Poynting vector of the directional scattering.}
\label{fig_scatter}
\end{figure}

\begin{figure}[htb]
\centering
{\includegraphics[width=7.8cm]{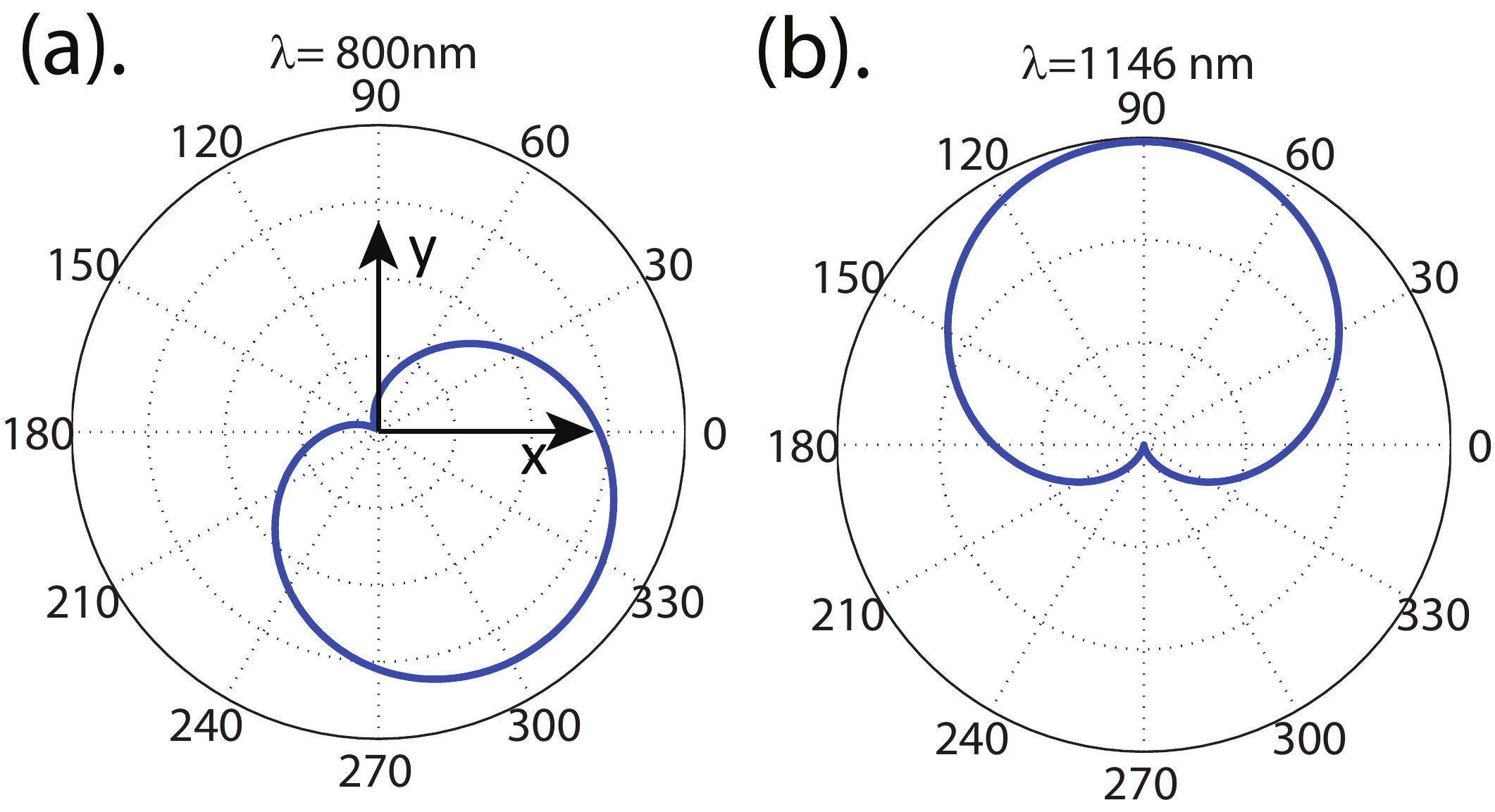}}
\caption{Scattering pattern at different wavelengths when $120\pi|b_1\mathbf{H}_f|=|a_1\mathbf{E}_f|$ are fullfilled for each wavelength, phase of the pupil field $\mathbf{E}_e$ is set to be $\phi_0=0$ for all the wavengths. (a).$\lambda=800$ nm; (b). $\lambda=1146$ nm.}
\label{fig_wavelength}
\end{figure}

As is shown in the remaining of the paper, the scattering direction of a high index dielectric nanosphere can be controlled by focusing an appropriate pupil field. Based on Mie's solution, the high-order multipole modes(including magnetic quadrupole) of a high index dielectric nanosphere are generally of narrow bandwidth. Electric and magnetic dipoles coexist and dominate for longer wavelengths, where the higher order modes are neglegible. In this paper, we will focus on this wavelength range. In order to realize directional scattering with the proposed concept, one must excite the circularly polarized magnetic dipole in the $xy$ plane and the electric dipole in $z$ direction and one must be able to control the amplitude and phase of these two dipoles separately. This cannot be achieved with conventional plane wave illumination. If one follows the idea in \cite{Xi_OL2016} for selective exciation of magnetic and electric dipoles, three focusing system will be needed: one along the $z$ axis to focus a radially polarized beam to excite the electric dipole, and two systems along the $x$ and $y$ axis, respectively, both with focused azimuthally polarized beams but $\pi/2$ phase difference to excite the circularly polarized magnetic dipole in the $xy$ plane. This is, however, unnecessarily complex.\\
In the current paper, we utilize the unique focusing properties of designed pupil fields with unconventional polarization and phase distributions to independently generate a circularly polarized magnetic field and a linearly polarized electric field at the focal point.\\
\begin{figure}[!h]
\centering
{\includegraphics[width=7.8cm]{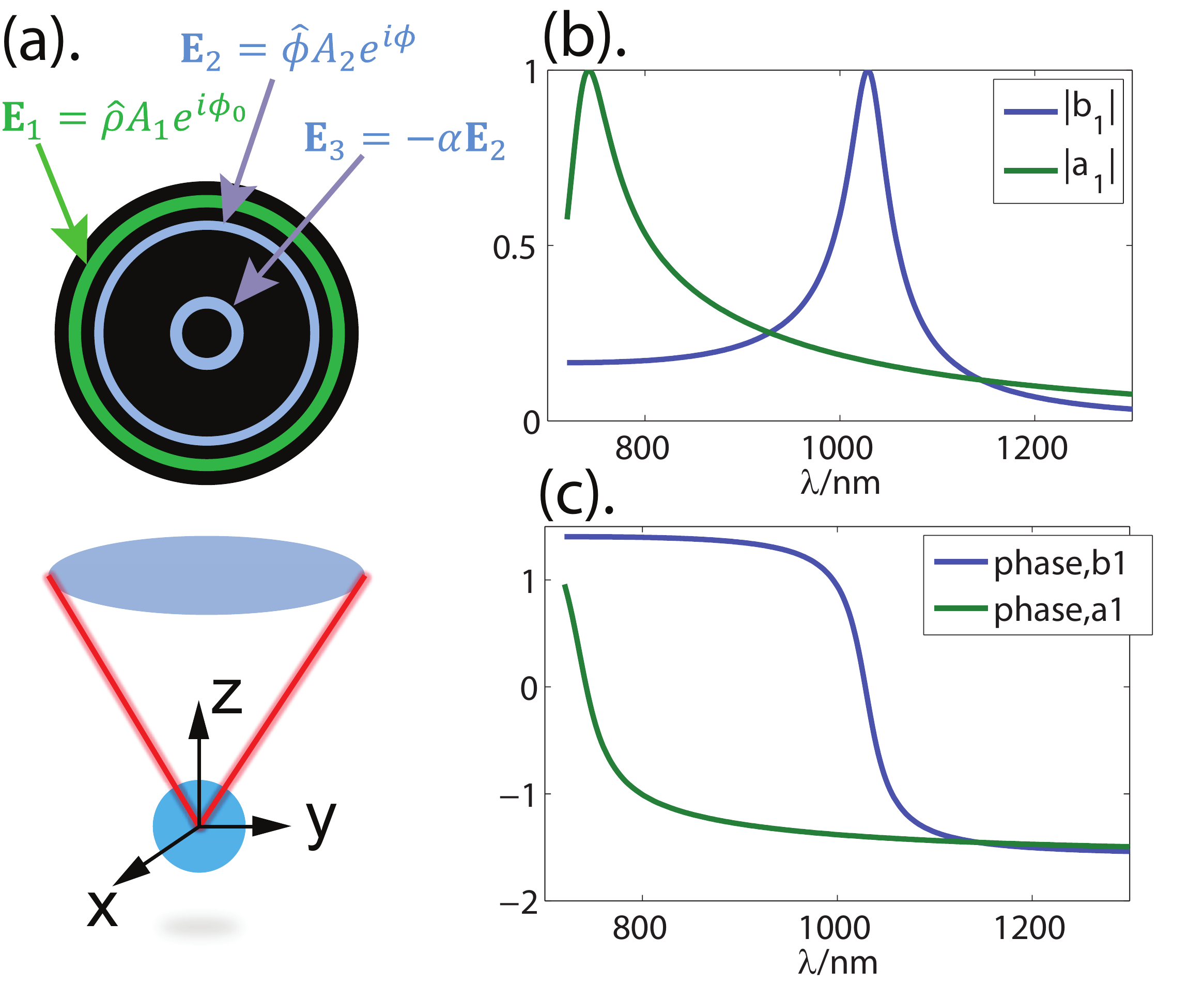}}
\caption{(a).Illustration of the configuration of focusing the pupil field to excite a high index dielectric sphere at focus; (b). Amplitude of the electric and magnetic scattering coefficients $a_1$ and $b_1$ as function of the wavelength for a high index dielectric sphere of refractive index $n=5$ and radius $r_0=100$ nm; (c).  Phase of the electric and magnetic scattering coefficients $a_1$ and $b_1$ of a high index dielectric sphere of refractive index $n=5$ and radius $r_0=100$ nm.}
\label{fig_configuration}
\end{figure}
As illustrated in Fig.\ref{fig_configuration}(a), the proposed pupil field contains three annular rings in the amplitude function. The radius of the pupil is normalized to 1. Within each ring, the amplitude is uniform. For simplicity, all three rings are chosen to have narrow width. The electrical field in ring 1 with average radius $\rho_1$ and width $\Delta\rho_1\ll1$ is radially polarized: $\mathbf{E}_{e}=\mathbf{E}_1=\hat{\rho}A_1e^{i\phi_0}$, where $\phi_0$ is a constant phase and $A_1$ is a positive real number.  Corresponding focused field can be calculated using the Richard Wolf diffraction integral\cite{TGBrown00,EWolf1958,Wei_JOptics}. In the focal point $r_f=0$ the focused electric field is longitudinally polarized, i.e. it is polarized in the $z-$ direction and:
\begin{eqnarray}
&E_{f,z}^{rad}(r_{f}=0)=-2e^{i\phi_0}C_0\int_0^1 A(\rho)\frac{s_0\rho}{(1-\rho^2s_0^2)^{1/4}}\rho \mathrm{d}\rho\\ \nonumber
&\approx -2e^{i\phi_0}C_0A_1\frac{s_0\rho_1}{(1-\rho_1^2s_0^2)^{1/4}}\rho_1\Delta\rho_1\\\nonumber
&=E_f e^{i\phi_0},
\label{eq0}
\end{eqnarray} 
where $s_0$ is the numerical aperture and $C_0$ is a constant related to the wavelength, numerical aperture and focal length. This pupil field is used to excite the pure electric dipole of the dielectric nanosphere. Note that for a radially polarized electric pupil field, the magnetic pupil field is azimuthally polarized and hence the magnetic field in the focal point vanishes. Therefore the pupil field $\mathbf{E}_{e}$ does not excite a magnetic dipole.\\
The pupil field $\mathbf{E}_h=\mathbf{E}_2+\mathbf{E}_3$ consists of two rings: ring 2 with average radius $\rho_2$ and width $\Delta\rho_2\ll1$ and ring 3 with average radius $\rho_3$ and width $\Delta\rho_3\ll1$, where $\rho_3\ll\rho_2<\rho_1$. The electric pupil fields in ring 2 and 3 are azimuthally polarized phase vortices: $\mathbf{E}_2=\hat{\phi}A_2e^{i\phi}$ and $\mathbf{E}_3=-\alpha\mathbf{E}_2$, where $\alpha$ and $A_2$ are positive real numbers. The corresponding magnetic pupil field is radially polarized with $\mathbf{H}_h=\hat{\rho}/Z_0|\mathbf{E}_h|e^{i\phi}$, where  $Z_0=120\pi$. Hence the magnetic pupil field is a radially polarized phase vortex. The corresponding field at the focal point $r_f=0$ can be calculated again using Richard Wolf diffraction integral\cite{TGBrown00,EWolf1958,Wei_JOptics}:
\begin{eqnarray}\label{eq1azm} 
&H_{f,x}^{azm,m=1}=\frac{C_0}{Z_0}\int_0^1 A(\rho)(1-\rho^2s_0^2)^{1/4}\rho \mathrm{d}\rho\\\nonumber
&\approx \frac{C_0}{Z_0}[A_2(1-\rho_2^2s_0^2)^{1/4}\rho_2\Delta\rho_2-\\\nonumber
&-\alpha A_2(1-\rho_3^2s_0^2)^{1/4}\rho_3\Delta\rho_3]=H_f,
\end{eqnarray}  
\begin{equation}
H_{f,y}^{azm,m=1}=iH_{f,x}^{azm,m=1}(r_f=0)=iH_f,
\label{eq2azm}
\end{equation} 
\begin{equation}
H_{f,z}^{azm,m=1}=0,
\label{eq3}
\end{equation} 
\begin{eqnarray} 
&E_{f,x}^{azm,m=1} = iC_0\int_0^1 A(\rho)(1-\rho^2s_0^2)^{-1/4}\rho \mathrm{d}\rho\\\nonumber
&\approx iC_0\times\\\nonumber
&\times[A_2(1-\rho_2^2s_0^2)^{-1/4}\rho_2\Delta\rho_2-\alpha A_2(1-\rho_3^2s_0^2)^{-1/4}\rho_3\Delta\rho_3],
\label{eq4}
\end{eqnarray}  
\begin{equation}\label{eq5}
E_{f,y}^{azm,m=1}=iE_{f,x}^{azm,m=1},
\end{equation} 
\begin{equation}
E_{f,z}^{azm,m=1}=0,
\label{eq6}
\end{equation} 
It follows from Eq. (\ref{eq1azm}) to Eq. (\ref{eq6}) that the transverse componets of the focal magnetic field have $(1-\rho^2s_0^2)^{1/4}$ dependence in the diffraction integral while  the transverse componet of the focal electric field have $(1-\rho^2s_0^2)^{-1/4}$ dependence. When
\begin{equation}
 \alpha=[(1-\rho_2^2s_0^2)^{-1/4}\rho_2\Delta\rho_2]/[(1-\rho_3^2s_0^2)^{-1/4}\rho_3\Delta\rho_3],
\end{equation}
all the electrical field components become zero at the focal point, but the magnetic field $\mathbf{H}_f=H_f(\hat{x}+i\hat{y})$ is non-zero and circularly polarized, where
\begin{eqnarray}
H_f&=\frac{C_0}{Z_0}A_2(1-\rho_2^2s_0^2)^{-1/4}\rho_2\Delta\rho_2\times\\\nonumber
&\times\left[\sqrt{1-\rho_2^2s_0^2}-\sqrt{1-\rho_3^2s_0^2}\right]\ne0.
\end{eqnarray}
In this way, a pure circularly polarized magnetic field at the focal point is created and the electric field vanishes there. \\
Note that the condition for $\alpha$ is independent of the wavelength. We remark that many other pupil fields can be designed to excite the electric and magnetic dipoles in the focal point. Indeed, the rings do not need to be narrow and by using broader rings, stronger dipole moments can be realized. However, the pupil field proposed in this paper have the advantage of being particularly simple.\\
It is shown in \cite{Das_PRB2015, AGE2011} that electric dipole and magnetic dipole interactions of a given dielectric nanosphere illuminated by an inhomogeneous beam depend only on the local fields at the center of the sphere. When a high index dielectric nanosphere at focus is illuminated by the focal field of the pupil field $\mathbf{E}=\mathbf{E}_h+\mathbf{E}_e$, then within the wavelength range where the magnetic and electric dipole modes of the sphere dominate, a circularly polarized magnetic dipole $\mathbf{m}=i\frac{6\pi}{k^3} b_1H_f (\hat{x}+i\hat{y})$  is excited with the pure contribution of pupil field $\mathbf{E}_h$ while the linearly polarized electric dipole $\mathbf{p}=i\frac{6\pi}{k^3}\epsilon_0 a_1E_f e^{i\phi_0}\hat{z}$ is excited with the pure contribution of pupil field $\mathbf{E}_e$. Here $a_1$ and $b_1$ are the electric and magnetic dipole scattering coefficients of a nanosphere. Applying Eq.\ref{eq_ed} and Eq.\ref{eq_md}, the total scattering field in the $xy$ plane of the excited magnetic and electric dipole moments is:
\begin{equation}
\mathbf{E}_s=\hat{z}\frac{6\pi}{k}(ia_1E_f e^{i\phi_0}+b_1Z_0 H_f e^{i\phi})\frac{e^{ikr}}{4\pi r}.
\label{eq_es}
\end{equation}
This shows that when the focal fields follow the relation $120\pi|b_1\mathbf{H}_f|=|a_1\mathbf{E}_f|$ which is the amplitude requirement of the Kerker's condition for directional scattering\cite{FU_NC2013}, there is always a direction in the $xy$ plane for which the phase condition for destructive interference $\phi=-\pi/2+\phi_0+Arg(a_1)-Arg(b_1)$  is fullfilled. \\
 In this paper, the physical model for arbitrary scattering direction control is verified by a finite element method(FEM) simulation\cite{wei2007} with which we calculated the scattering of a nano-sphere under the proposed focused beam excitation. A dielectric sphere of refractive index 5 and radius 100 nm is considered but the same principle applies to nanospheres with different high indices and different radius. The electric and magnetic dipole scattering coefficients $a_1$ and $b_1$  of the nanosphere are shown in Fig.\ref{fig_configuration}(b) and (c). When the amplitudes of the pupil field $\mathbf{E}_e$ and $\mathbf{E}_h$ are modulated in such a way that the focal fields satisfy the relation $120\pi|b_1\mathbf{H}_f|=|a_1\mathbf{E}_f|$, there is always a direction in the $xy$ plane for which the phase condition for directional scattering is fullfilled due to the spiral phase of the scattering electric field of the spinning magnetic dipole as shown in Fig.\ref{fig_scatter}(a) and the cylindrical symmetry as depicted in  Fig.\ref{fig_scatter}(b) of the scattering field of the electric dipole. As is shown in Fig.\ref{fig_scatter}(c), when the amplitude requirement of Kerker's condition is satisfied for $\lambda=1028$ nm, the total electrical field is formed by the interference of these two scattering fields. As a result, a destructive interference is formed along the direction $\phi=-\pi/2+\phi_0+Arg(a_1)-Arg(b_1)$  in the $xy$ plane, and the directional scattering is thus achieved, clearly illustrated by the plot of the Poynting vector in Fig.\ref{fig_scatter}(d).\\
\begin{figure}[htb]
\centering
{\includegraphics[width=7.8cm]{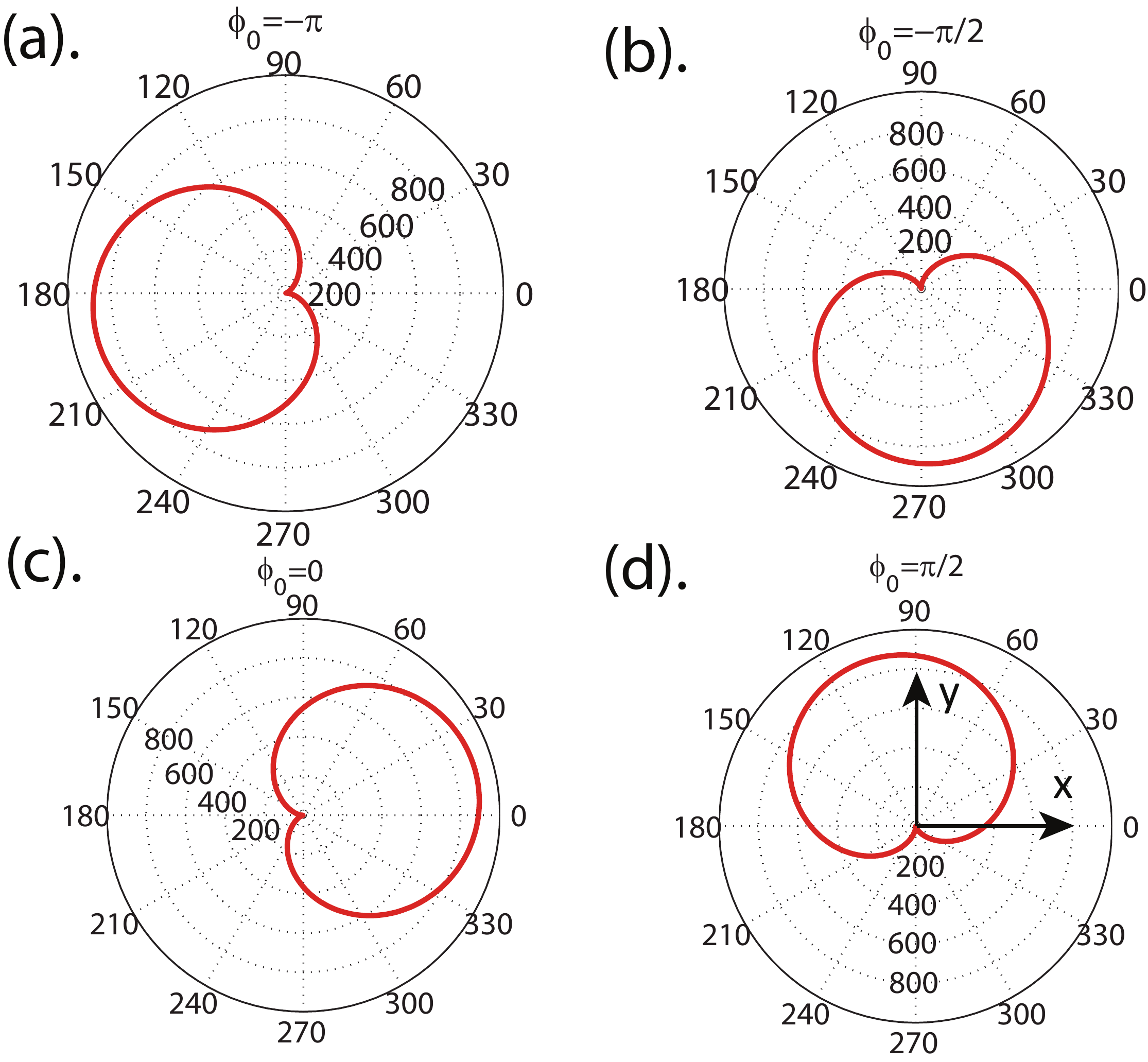}}
\caption{Scattering pattern changes when $\phi_0$ is changed over $2\pi$ with illumination wavelength $\lambda=1028$ nm. (a).$\phi_0=-\pi$; (b).$\phi_0=-\pi/2$; (c). $\phi_0=0$; (d). $\phi_0=\pi/2$.}
\label{fig_phi0}
\end{figure}
As shown in Fig.\ref{fig_phi0}, by changing the phase $\phi_0$ of the pupil field of $\mathbf{E}_e=\hat{\rho}A_1 e^{i\phi_0}$, the scattering can be tuned. When $\phi_0$ is modified over $2\pi$ range with a $\pi/2$ step, the scattering direction also changes over $2\pi$ range, making it possible to tune the scattering to any direction in the $xy$ plane by only modifying the phase of the radially polarized beam.\\
Furthermore, this approach to control directional scattering is broadband. As long as the wavelength and the radius of the sphere are such that the electric and magnetic dipole modes dominate, the requirement for the focal field $120\pi|b_1\mathbf{H}_f|=|a_1\mathbf{E}_f|$ is fullfilled by adjusting the amplitude of the pupil field $\mathbf{E}_e$ and $\mathbf{E}_h$.  Then, the directional scattering is always achieved as shown in Fig.\ref{fig_wavelength}. \\
It is important to note that although we focus on the interference of a spinning magnetic dipole and a linearly polarized electric dipole, the directional scattering control can also be achieved with a spinning electric dipole and a linearly polarized magnetic dipole. The corresponding excitation focal field can be realized by applying an azimuthally polarized electric pupil field in ring 1 and adjusting $\alpha$ in ring 3 so that the magnetic field at focus is zero but the electric field is nonzero and circularly polarized.\\
\section{Conclusion}
In summary, we have proposed an approach to control the scattering direction of a high index dielectric nanosphere. This approach enables independent excitation of a circularly polarized magnetic and a linearly polarized electric dipole mode with the unique focusing properties of designed pupil with unconventional polarization and phase distributions. By simply adjusting the phase and amplitude of the pupil field, the scattering can be directed in any direction in a plane perpendicular to the optical axis of the focused field. Moreover, the directional scattering is achieved over a relatively broad wavelength range. Our method has high potential in many application such as control of the light routing on a photonic chip, selective excitation of fluorescent molecules for superresolution microscopy, optical switching, etc.\\

\end{document}